\documentclass[useAMS,usenatbib]{mn2e}
\usepackage{graphicx}
\usepackage{txfonts}
\usepackage{color} 

\title[Abundance gradients in irregulars]
      {On the radial abundance gradients in disks of irregular galaxies} 
      
\author[L.S.~Pilyugin et al.]
       {L.S.~Pilyugin$^{1,2,3}$,
        E.K.~Grebel$^{1}$,
        I.A.~Zinchenko$^{1,2}$  \\
       $^{1}$ Astronomisches Rechen-Institut, Zentrum f\"{u}r Astronomie 
             der Universit\"{a}t Heidelberg, 
             M\"{o}nchhofstr.\ 12--14, 69120 Heidelberg, Germany \\
       $^{2}$ Main Astronomical Observatory
             of National Academy of Sciences of Ukraine,
             27 Zabolotnogo str., 03680 Kiev, Ukraine \\   
       $^{3}$ Kazan Federal University, 18 Kremlyovskaya St., 420008, Kazan. Russian Federation
        }

\date{Accepted 2015 April 22. Received 2015 April 22; in original form 2015 January 01}

\begin{document}

\maketitle

\begin{abstract}
We determine the radial abundance distributions across the disks of
fourteen irregular galaxies of the types {\em Sm} and {\em Im}
(morphological $T$ types $T$ = 9 and $T$ =10) as traced by their
H\,{\sc ii} regions.  The oxygen and nitrogen abundances in
H\,{\sc ii} regions are estimated through the $T_{e}$ method or/and
with the counterpart method ($C$ method). Moreover, we examine the
correspondence between the radial abundance gradient and the surface
brightness profile.  We find that irregular galaxies with a flat inner
profile (flat or outwardly increasing surface brightness in the
central region) show shallow (if any) radial abundance gradients.  On
the other hand, irregular galaxies with a steep inner profile (with or
without a bulge or central star cluster) usually show rather steep
radial abundance gradients. This is in contrast to the widely held
belief that irregular galaxies do not usually show a radial abundance
gradient. 
\end{abstract}


\section{Introduction}

The radial distribution of gas-phase oxygen abundances traced by
H\,{\sc ii} regions has been investigated in the disks of many spiral
galaxies
\citep{VilaCostas1992,Zaritsky1994,vanZee1998,Pilyugin2004,Moustakas2010,Gusev2012,Pilyugin2014a,Sanchez2014}.
It was found that almost all spiral galaxies show radial abundance
gradients in the sense that their inner H\,{\sc ii} regions (i.e.,
those closer to the galactic centers) have higher oxygen abundances
than the outer ones.  

The radial distribution of abundances across the disks of irregular
galaxies is less well studied.  \citet{Pagel1978} analyzed spectra of
a number of  H\,{\sc ii} regions in the Small and Large Magellanic
Clouds. They determined the abundances in H\,{\sc ii} regions through
the effective temperature ($T_{e}$) method using their own
measurements together with the spectral measurements by other authors
and examined the spatial distributions of abundances in those
galaxies.  \citet{Pagel1978} concluded that any radial abundance
gradient in present-day abundances is small or absent in the Large
Magellanic Cloud and is conspicuously absent in the Small Magellanic
Cloud.  In the Small Magellanic Cloud, also stellar metallicity
determinations support the absence of a radial gradient, although
there is a large metallicity spread of $\sim 0.6$ dex in [Fe/H] for
a given age \citep{Glatt2008,Cignoni2013}. 

\citet{Roy1996} studied the oxygen abundance distributions in the
disks of the dwarf irregular galaxy NGC~2366 and the dwarf Seyfert I
galaxy NGC~4395 using imaging spectrophotometry with narrow-band
filters in the lines of H$\alpha$, H$\beta$, [O\,{\sc
iii}]$\lambda$5007 and [N\,{\sc ii}]$\lambda$6584.  They used the line
ratio [O\,{\sc iii}]/[N\,{\sc ii}] as an abundance indicator (O3N2
calibration).  They found that there is no global oxygen abundance
gradient across the disks of those galaxies. 

\citet{Hunter1999} obtained emission-line long-slit spectra of 189
H\,{\sc ii} regions in a sample of 65 {\em Im}, {\em Sm}, and blue
compact dwarf galaxies. They estimated the oxygen abundances in
H\,{\sc ii} regions using the line ratio [O\,{\sc iii}]/[N\,{\sc ii}]
(O3N2 calibration) and the combination of $R_{23}$ = [O\,{\sc
iii}]+[O\,{\sc ii}] and [O\,{\sc iii}]/[O\,{\sc ii}] (two-dimensional
$R_{23}$ calibration) when the oxygen line [O\,{\sc ii}]$\lambda$3727
was measured.  \citet{Hunter1999} examined the radial abundance
distribution in disks of eight {\em Sm} and {\em Im} galaxies for
which they measured at least three H\,{\sc ii} regions.  They found
that the oxygen abundances within a given galaxy generally vary by
about 0.2 dex, but they did not detect a trend in oxygen abundances
with radius except for the {\em Sm} galaxy DDO~204.

\citet{Kniazev2005} measured oxygen abundances with the direct method
in three H\,{\sc ii} regions in each of the dwarf irregulars Sextans~A
and Sextans~B.  While they found Sex~A to be chemically homogeneous,
one of the three H\,{\sc ii} regions in Sex~B turned out to be about
twice as metal-rich than the other two, and the abundances of other
heavy elements suggest an enrichment by a factor of $\sim 2.5$ as
compared to the other two H\,{\sc ii} regions.  \citet{Kniazev2005}
attribute this to inhomogeneous chemical enrichment. 

\citet{vanZee2006} carried out long-slit spectroscopy of 67 H\,{\sc
ii} regions in 21 dwarf irregular galaxies. Oxygen abundances for 25
H\,{\sc ii} regions were derived through the direct $T_{e}$ method;
the abundances in other  H\,{\sc ii} regions were estimated using
strong line calibrations. \citet{vanZee2006} considered the oxygen
abundances as a function of radius for 12 irregular galaxies with
three or more observations and found that the abundances are very
similar (within the formal errors) within each galaxy with the
possible exception of the galaxy UGC~12894.  \citet{vanZee2006} noted
that the radial trend in oxygen abundances (three points) in the
UGC~12894 may be artificial because the abundances of the inner and
outer H\,{\sc ii} regions were obtained via different methods (through
the strong line calibration for two inner H\,{\sc ii} regions and
through the $T_{e}$ method for outer H\,{\sc ii} region). 

Similarly, \citet{Lee2007} obtained oxygen abundances for 35  H\,{\sc
ii} regions in eight dwarf galaxies in the Centaurus~A group and in 13
H\,{\sc ii} regions in closer dwarfs.  Some of their measurements use
the direct $T_{e}$ method, while the majority of the abundance
determinations is based on strong-line calibrations.  Although the
results for individual H\,{\sc ii} regions in a given galaxy tend to
vary, Lee et al.\  point out that the variations are within the
uncertainties of the strong-line method.  In one of the dwarf
irregulars of the Cen~A group, AM~$1318-444$, one of the H\,{\sc ii}
regions is considerably more oxygen-rich than the others.  Lee et al.\
argue that the measured line intensity ratios suggest that this
emission nebula is a supernova remnant.  They also note that radial
gradients may exist in some of their targets such as in the {\em Sm}
NGC\,3109 or NGC\,5264, but that more and deeper data are needed to
establish this. 

It is the current belief that irregular galaxies generally do not show
radial abundance gradients in their young populations and are
chemically homogeneous.  This implies that there is a ``spiral versus
irregular dichotomy" in the sense that there is a sudden change from
spiral (radial abundance gradients are usually present) to irregular
galaxies (typically no gradients).  However, other properties (e.g.,
gas fraction, global metallicity) vary smoothly in transition from
spirals to irregulars \citep[][among
others]{Zaritsky1994,Pilyugin2000,Garnett2002,Pilyugin2007}.

The measurements of the abundance gradients in the disks of irregular
galaxies often encounter the following difficulty.  Reliable oxygen
abundances in a number of H\,{\sc ii} regions in the disk of a galaxy
should be determined in order to evaluate the existence of an
abundance gradient.  Abundance determinations using the direct
$T_{e}$ method require high-precision spectroscopy including the weak
auroral lines [O\,{\sc iii}]$\lambda$4363 or/and [N\,{\sc
ii}]$\lambda$5755.  Unfortunately, these weak auroral lines are
usually only detected in the spectra of a few (if any) of the
brightest H\,{\sc ii} regions in a given irregular galaxy. The oxygen
abundances in the other H\,{\sc ii} regions are then estimated using
the strong-line method pionered by \citet{Pagel1979} and
\citet{Alloin1979}.  The principal idea of the strong-line method is
to establish the relation between the (oxygen) abundance in an H\,{\sc
ii} region and some combination of the intensities of strong emission
lines in its spectrum (such a relation is usually called a
``calibration'').  Different calibrations were suggested. A prominent
characteristic of the calibrations is that they are not applicable
across the whole range of metallicities of H\,{\sc ii} regions but
only within a limited interval (usually only at high or at low
metallicities).  The oxygen abundances of irregular galaxies typically
lie within or near the transition zone in the $R_{23}$ -- O/H diagram
(from 12 + log(O/H) $\sim$7.9 to $\sim$8.3) where calibrations cannot
be used or where they provide abundances with large uncertainties. 

We recently suggested a new method (the ``C method'') for abundance
determinations in  H\,{\sc ii} regions, which can be used over the
whole range of metallicities of H\,{\sc ii} regions and which provides
oxygen and nitrogen abundances on the same metallicity scale as the
classic $T_{e}$ method \citep{Pilyugin2012,Pilyugin2013}. Using this
method, we examined the abundance gradients in the disks of 130
late-type galaxies including several irregular galaxies
\citep{Pilyugin2014a}.  In that study, radial abundance gradients were
found in irregular galaxies. Here we will focus on the investigation
of the abundance gradients in a sample of irregular galaxies ({\em Sm}
and {\em Im}, morphological $T$ types 9 and 10). Since there is a
relation between oxygen abundance and disk surface brightness in
spiral galaxies \citep[e.g.,][]{Pilyugin2014b}, we will also examine
the relation between radial abundance distributions and surface
brightness profiles of the disks of irregular galaxies. 

The paper is structured as follows.  The spectral and photometric data
are reported in Section 2.  The radial abundance gradients are
determined in Section 3.  The discussion and conclusions are given in
Section 4, followed by a summary (Section 5).

\section{The data}

\subsection{Our sample}

We have selected a sample of irregular {\em Sm} and {\em Im} and
galaxies with morphological $T$ types of 9 and 10 according to the
RC3 catalog \citep{RC3}. It should be noted that the morphological
classification of some galaxies is not robust. The morphological $T$
types in different sources can differ by up to 1. We only consider
irregular galaxies with available spectra for four and more  H\,{\sc
ii} regions.  The validity of the radial abundance is defined not only
by the quantity and quality of the spectra but also by the
distribution of the measured  H\,{\sc ii} regions along the galactic
radius. We reject galaxies where the measured  H\,{\sc ii} regions
cover less than $\sim$~1/3 of the optical radius of a galaxy. For
example, for this reason we rejected the galaxy UGC~5666 (a.k.a.\
IC~2574 or DDO~81). More than ten spectra are available for this
galaxy \citep{Miller1996,Croxall2009}, but the measured H\,{\sc ii}
regions cover only a small fraction of the optical radius of the
galaxy, which prevents a reliable investigation of a radial
abundance gradient. 

Our final list includes fourteen irregular galaxies with optical radii
of $R_{25}$ $\ga$ 2 kpc.  Table \ref{table:sample} lists the general
characteristics of each galaxy.  The column 1 contains the order
number.  The columns 2 -- 4  give the galaxy's name. We list the
number of a galaxy acording to the New General Catalogue (NGC, column
2), the Uppsala General Catalog of Galaxies (UGC, column 3), and one
other widely used name (column 4).  The morphological type of the
galaxy (morphological type code $T$) from the RC3 is reported in
column 5.  The right ascension (R.A.) and declination (Dec.) (J2000.0)
of each galaxy are given in columns 6 and 7.  The right ascension and
declination are obtained from our photometry (see Section 2.3) or
taken from the NASA/IPAC Extragalactic Database ({\sc
ned})\footnote{The NASA/IPAC Extragalactic Database ({\sc ned}) is
operated by the Jet Populsion Laboratory, California Institute of
Technology, under contract with the National Aeronautics and Space
Administration.  {\tt http://ned.ipac.caltech.edu/} }.  The position
angle (P.A.), axis ratio ($b/a$), and inclination are listed in
columns 8 -- 10.  The isophotal radius $R_{25}$ in arcmin and in kpc
of each galaxy is reported in columns 11 and 12, respectively.  The
adopted distance $d$ taken from \citet{Karachentsev2013} or from the
{\sc ned} is reported in column 13.  The {\sc ned} distances use flow
corrections for Virgo, the Great Attractor, and Shapley Supercluster
infall.  The references to sources for geometrical parameters (first
reference) and for distances (second reference) are given in column
14.

\begin{table*}
\caption[]{\label{table:sample}
The adopted properties of our galaxies.
}
\begin{center}
\begin{tabular}{rrrlcccrcccccl} \hline \hline
$n$                          &
\multicolumn{3}{c}{Name}     &
$T$                          &
R.A.                         &
Dec.                         & 
P.A.                         &
$b/a$                        &
Inclination                  &
R$_{25}$                      &
R$_{25}$                      &
$d$                           &
Reference                    \\
                             &
NGC                          &
UGC                          &  
Other                        &
type                         &
                             &
                             &
                             &
                             &
degree                       &
arcmin                       &
kpc                          &
Mpc                          &
                             \\  \hline 
  1 &         &    2023  & DDO 25         &  10  & 02:33:18.20  &   33:29:28.0  &       &  1.00 &   0  &  0.83  &   2.24  &    9.30  &  RC3;  K13$^{a}$  \\
  2 &         &    2216  &                &  10  & 02:44:21.14  &   00:40:37.6  &   13  &  0.41 &  66  &  0.40  &   4.25  &   36.50  &  Here; {\sc ned} \\
  3 &  1156   &    2455  &                &  10  & 02:59:42.30  &   25:14:16.2  &   25  &  0.74 &  42  &  1.66  &   3.76  &    7.80  &  RC3;  K13       \\
  4 &  2537   &    4274  &                &   9  & 08:13:14.66  &   45:59:32.7  &    5  &  0.95 &  18  &  1.07  &   3.80  &   12.20  &  Here; K13       \\
  5 &         &    4305  & DDO 50         &  10  & 08:19:04.98  &   70:43:12.1  &   15  &  0.79 &  37  &  3.97  &   3.92  &    3.39  &  RC3;  K13       \\
  6 &  3738   &    6565  &                &  10  & 11:35:48.81  &   54:31:25.8  &  167  &  0.66 &  49  &  1.43  &   2.04  &    4.90  &  Here; K13       \\
  7 &         &    6980  &                &  10  & 11:59:06.20  &   24:28:20.3  &  159  &  0.48 &  61  &  0.23  &   3.55  &   53.10  &  Here; {\sc ned} \\
  8 &  4214   &    7278  & NGC 4228       &  10  & 12:15:39.19  &   36:19:36.6  &  126  &  0.89 &  27  &  3.18  &   2.72  &    2.94  &  Here; K13       \\
  9 &  4395   &    7524  &                &   9  & 12:25:48.88  &   33:32:48.7  &  135  &  0.64 &  50  &  4.08  &   5.47  &    4.61  &  Here; K13       \\
 10 &         &    7557  &                &   9  & 12:27:11.24  &   07:15:47.1  &  148  &  0.80 &  37  &  1.14  &   4.54  &   13.70  &  Here; {\sc ned} \\
 11 &  4449   &    7592  &                &  10  & 12:28:11.01  &   44:05:38.1  &   48  &  0.56 &  56  &  3.07  &   3.76  &    4.21  &  Here; K13       \\
 12 &         &          & CGCG 071-090   &  10  & 12:58:52.80  &   13:09:08.8  &  172  &  0.60 &  53  &  0.49  &   1.98  &   13.90  &  Here; {\sc ned} \\
 13 &         &    9614  &                &  10  & 14:56:47.70  &   09:30:33.4  &   21  &  0.77 &  40  &  0.50  &   7.23  &   49.70  &  Here; {\sc ned} \\
 14 &         &   12709  & DDO 219        &   9  & 23:37:24.05  &   00:23:30.7  &  150  &  0.73 &  43  &  0.73  &   7.86  &   37.00  &  Here; {\sc ned} \\
\hline
\end{tabular}\\
\end{center}
\begin{flushleft}
$^a$  K13 -- \citet{Karachentsev2013} 
\end{flushleft}
\end{table*}

\subsection{Emission line intensities in the H\,{\sc ii} region spectra}

We use the  emission line intensities in published spectra of  H\,{\sc
ii} regions from different works for abundance determinations. We have
searched for spectra of H\,{\sc ii} regions with measured H$\alpha$,
H$\beta$, [O\,{\sc iii}]$\lambda$5007, [N\,{\sc ii}]$\lambda$6584, and
[S\,{\sc ii}]$\lambda$6717+$\lambda$6731 lines. The lines [O\,{\sc
ii}]$\lambda$3727+$\lambda$3729 and  [O\,{\sc iii}]$\lambda$4363 are
also available in a number of spectra. We have taken the de-reddened
line intensities as reported by the authors. If only the measured
fluxes are given then the measured emission-line fluxes were corrected
for interstellar reddening in the same way as in
\citet{Pilyugin2014a}.  These spectroscopic data form the basis for
the abundance determinations. The references to the spectroscopic data
sources are listed in Table~\ref{table:grad}.

\subsection{Surface brightness profiles}

\begin{figure}
\resizebox{1.00\hsize}{!}{\includegraphics[angle=000]{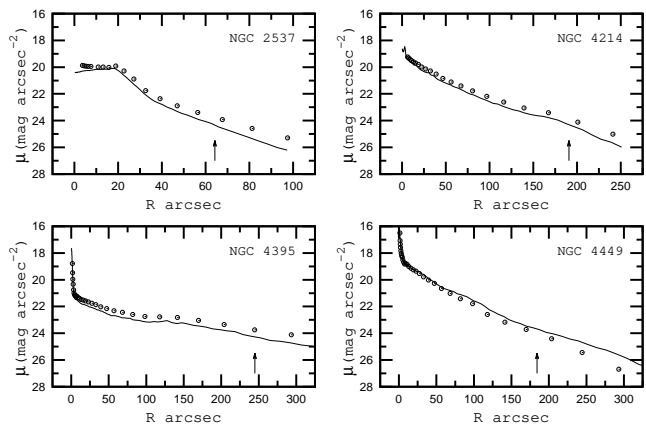}}
\caption{ Comparison between the measured surface brightness profiles 
in the SDSS $r$ band (line) obtained here and profiles in the $R$ band 
(open circles) from \citet{Swaters2002}. The arrow marks the optical 
isophotal radius $R_{25}$. }
\label{figure:igor-swat}
\end{figure}

We constructed radial surface brightness profiles in the infrared $W1$
band (with an isophotal wavelength of 3.4 $\mu$m) using the publicly
available photometric maps obtained in the framework of the {\it
Wide-field Infrared Survey Explorer (WISE)} project
\citep{Wright2010}.  The conversion of the photometric map into the
surface brightness profile is discussed in \citet{Pilyugin2014b}.
Parameters such as the galaxy center, the position angle of the
major axis, and the axis ratio are obtained through fitting of the
isophotes by ellipses.

We also constructed radial surface brightness profiles in the SDSS $g$
and $r$ bands using the photometric maps of SDSS data release 9
\citep{Ahn2012ApJS203}. To estimate the optical isophotal radius
$R_{25}$ of a galaxy, the surface brightnesses in the SDSS filters $g$
and $r$ were converted to $B$-band brightnesses, and the $AB$
magnitudes were reduced to the Vega photometric system using the
conversion relations and solar magnitudes of \citet{Blanton2007}.

\citet{Swaters2002} reported surface brightness profiles in the $R$
band for a large number of galaxies.   Fig.~\ref{figure:igor-swat}
shows the comparison between our measured surface brightness profiles
in the SDSS $r$ band (solid line) and $R$-band profiles (open circles)
from \citet{Swaters2002}.  Our surface brightness profiles within the
optical isophotal radius $R_{25}$ agree satisfactorily well with those
of \citet{Swaters2002}. 

All surface brightness measurements were corrected for Galactic
foreground extinction using the $A_V$ values from the recalibration of
the maps of \citet{Schlegel1998} by \citet{Schlafly2011} and the
extinction curve of \citet{Cardelli1989}, assuming a ratio of total to
selective extinction of $R_{V}$ = $A_{V}$/$E_{B-V}$ = 3.1. The $A_V$
values given in the NASA Extragalactic Database ({\sc ned}) were
adopted.  To transform the surface brightness measurements to solar
units, we used the magnitude of the Sun in the $W1$ band, which we
obtained from its magnitude in the $V$ band and from its color
$(V-W1)_{\sun}$ = 1.608 taken from \citet{Casagrande2012}.  

The radial profiles in the SDSS $g$ and $r$ bands were used to
estimate the isophotal $R_{25}$ radius of each galaxy. The obtained
radial profiles were reduced to a face-on galaxy orientation. Note
that the inclination correction is purely geometrical, and it does not
include any correction for inclination-dependent internal obscuration.
The values of the optical radius $R_{25}$ determined here are listed
in Table~\ref{table:sample}. There are no SDSS photometric maps for
several galaxies of our sample. The optical radii $R_{25}$ (as well as
the position angle of the major axis and the inclination angle) for
those galaxies were taken from the RC3 \citep{RC3}. 

\begin{figure*}
\resizebox{1.00\hsize}{!}{\includegraphics[angle=000]{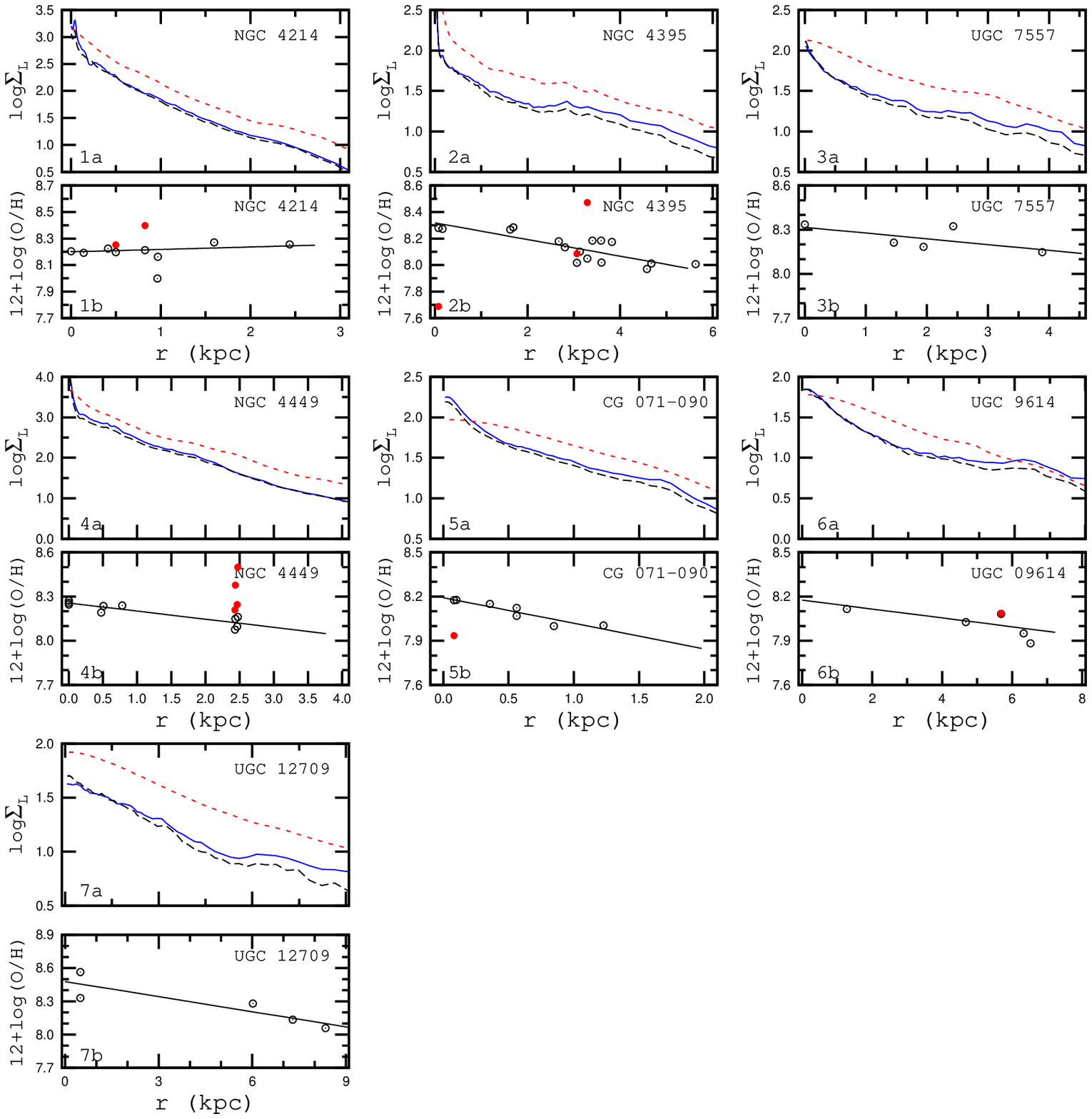}}
\caption{Surface brightness profiles and radial distributions of 
oxygen abundances for irregular galaxies with steep inner 
profiles. Each galaxy is presented in two panels. Each upper panel 
$N$a shows the surface brightness profiles in the SDSS $g$ band as 
a light-grey (blue) solid line, in the SDSS $r$ band as a dark 
(black) long-dashed line, and in the WISE $W1$ band as a dark-grey 
(red) short-dashed line.  
Each lower panel $N$b shows the oxygen abundance in individual 
H\,{\sc ii} regions as a function of radius.  The dark (black) open 
circles show (O/H)$_{C_{NS}}$ abundances and the grey (red) filled 
circles indicate the (O/H)$_{T_{e}}$ abundances. The solid line 
represents the inferred linear abundance gradient. 
(A color version of this figure is available in the on-line edition.)  }
\label{figure:ygrad}
\end{figure*}

The observed surface brightness profile of an irregular galaxy 
can be fitted by an exponential \citep{Swaters2002,Herrmann2013}.
There are bulges, bars, or nuclear star clusters at the centres of
some irregular galaxies. A bulge or a nuclear star cluster can be
fitted with a general S\'{e}rsic profile.  A profile showing a
increase of (optical) surface brightness in the central part of a
galaxy (with or without bulge-like component)  will be referred to as
a steep inner profile below. Irregular galaxies with a steep inner
profile are presented in Fig.~\ref{figure:ygrad}.

\begin{figure*}
\resizebox{1.00\hsize}{!}{\includegraphics[angle=000]{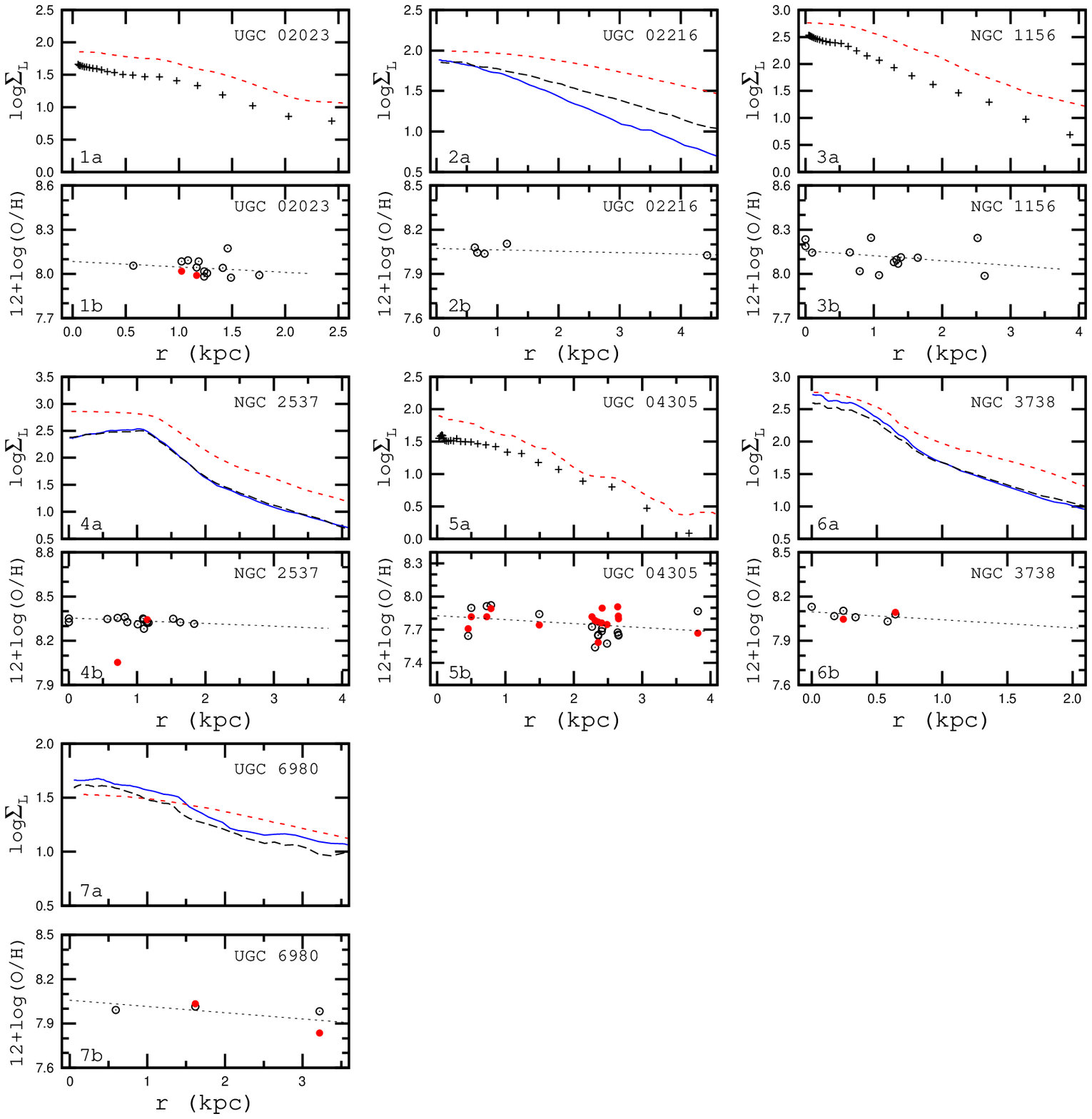}}
\caption{ The same as Fig.~\ref{figure:ygrad} but for irregular 
galaxies with flat inner profiles.  For galaxies without SDSS 
photometric maps, the surface brightness profile in the $R$ band 
from \citet{Swaters2002} is indicated with dark (black) plus signs.
(A color version of this figure is available in the on-line edition.)}
\label{figure:ngrad}
\end{figure*}

It is known that the surface brightnesses in some irregular galaxies
are flat or even increase out to a region of slope change where they
tend to fall off \citep{Swaters2002,Taylor2005,Herrmann2013}. 
Such surface brightness profiles can be formally fitted by an
exponential disk with a bulge-like component of negative brightness.
Such profiles will be referred to as flat inner profiles below. 
Irregular galaxies with flat inner profiles are presented in
Fig.~\ref{figure:ngrad}.

To define the type of surface brightness profile we use our
surface brightness profiles in the SDSS $r$ band or $R$-band profiles
from \citet{Swaters2002}.  It should be noted that the shapes of the
surface brightness profiles of the same galaxy in the different
photometric bands do not necessarily coincide with each other.

\section{Abundances}

\subsection{Abundance determination}

We determine the $T_{e}$-based oxygen (O/H)$_{T_{e}}$ and nitrogen
(N/H)$_{T_{e}}$ abundances in H\,{\sc ii} regions where the auroral
line [O\,{\sc iii}]$\lambda$4363 is detected using the equations of
the $T_{e}$-method from \citet{Pilyugin2010,Pilyugin2012}. 

A new method (called the ``$C$ method'') for oxygen and nitrogen
abundance determinations from strong emission lines has recently
been suggested \citep{Pilyugin2012,Pilyugin2013}.  Here, the strong
lines $R_3$ = [O\,{\sc iii}]$\lambda$$\lambda$4959,5007, $N_2$ =
[N\,{\sc ii}]$\lambda$$\lambda$6548,6584  and $S_2$ = [S\,{\sc
ii}]$\lambda$$\lambda$6717,6731 are used in the determinations of the
oxygen (O/H)$_{C_{\rm NS}}$ and nitrogen (N/H)$_{C_{\rm NS}}$
abundances in individual H\,{\sc ii} regions of our target galaxies.

\subsection{Radial abundance gradients}

\begin{table*}
\caption[]{\label{table:grad}
The derived parameters of the radial oxygen and nitrogen abundance 
distributions in our target galaxies.  
}
\begin{center}
\begin{tabular}{lrccccccl} \hline \hline
Galaxy                       &
$R_{25}$                      &
12+log(O/H)$_{R_{0}}$          &
O/H gradient                 &
$\sigma$(O/H)                &
12+log(N/H)$_{R_{0}}$          &
N/H gradient                 &
$\sigma$(N/H)                &
References                  \\
                             &
kpc                          &  
                             &
dex~R$_{25}^{-1}$              &
dex                          &
                             &
dex~R$_{25}^{-1}$              &
dex                          &    \\  \hline 
 UGC 02023      &  2.24  & 8.08 $\pm$ 0.07 & -0.082 $\pm$ 0.128 (0.150) & 0.052 & 6.72 $\pm$ 0.12 & -0.137 $\pm$ 0.208 (0.252) &  0.085  &  10, 18                  \\ 
 UGC 02216      &  4.25  & 8.07 $\pm$ 0.02 & -0.040 $\pm$ 0.043 (0.300) & 0.026 & 6.69 $\pm$ 0.05 & -0.056 $\pm$ 0.091 (0.560) &  0.054  &  16                      \\ 
 NGC 1156       &  3.76  & 8.16 $\pm$ 0.04 & -0.124 $\pm$ 0.112 (0.138) & 0.080 & 6.84 $\pm$ 0.07 & -0.232 $\pm$ 0.202 (0.260) &  0.145  &  8, 10, 14               \\ 
 NGC 2537       &  3.80  & 8.35 $\pm$ 0.01 & -0.071 $\pm$ 0.034 (0.031) & 0.014 & 7.28 $\pm$ 0.03 & -0.237 $\pm$ 0.113 (0.124) &  0.046  &  1, 6, 8, 16             \\ 
 UGC 04305      &  3.92  & 7.83 $\pm$ 0.04 & -0.140 $\pm$ 0.078 (0.085) & 0.101 & 6.31 $\pm$ 0.05 & -0.182 $\pm$ 0.090 (0.108) &  0.116  &  3, 4                    \\ 
 NGC 3738       &  2.04  & 8.10 $\pm$ 0.02 & -0.110 $\pm$ 0.103 (0.128) & 0.028 & 6.71 $\pm$ 0.04 & -0.144 $\pm$ 0.173 (0.201) &  0.046  &  1, 8, 10                \\ 
 UGC 06980      &  3.55  & 8.06 $\pm$ 0.07 & -0.150 $\pm$ 0.112 (0.124) & 0.056 & 6.62 $\pm$ 0.05 & -0.145 $\pm$ 0.077 (0.089) &  0.038  &  7, 16                   \\ 
 NGC 4214       &  2.72  & 8.20 $\pm$ 0.05 &  0.049 $\pm$ 0.124 (0.122) & 0.090 & 6.89 $\pm$ 0.06 &  0.037 $\pm$ 0.148 (0.189) &  0.108  &  8, 10, 11, 16           \\ 
 NGC 4395       &  5.47  & 8.32 $\pm$ 0.04 & -0.343 $\pm$ 0.072 (0.066) & 0.064 & 7.14 $\pm$ 0.07 & -0.629 $\pm$ 0.119 (0.085) &  0.104  &  5, 13, 16, 17           \\ 
 UGC 07557      &  4.54  & 8.32 $\pm$ 0.06 & -0.176 $\pm$ 0.118 (0.176) & 0.057 & 7.15 $\pm$ 0.14 & -0.361 $\pm$ 0.280 (0.413) &  0.135  &  15                      \\ 
 NGC 4449       &  3.76  & 8.26 $\pm$ 0.01 & -0.207 $\pm$ 0.034 (0.063) & 0.028 & 7.03 $\pm$ 0.03 & -0.369 $\pm$ 0.053 (0.122) &  0.051  &  1, 2, 8, 9, 12,         \\ 
                &        &                 &                            &       &                 &                            &         &                  13, 16  \\ 
 CG071-090      &  1.98  & 8.19 $\pm$ 0.02 & -0.344 $\pm$ 0.060 (0.074) & 0.026 & 6.86 $\pm$ 0.02 & -0.525 $\pm$ 0.074 (0.104) &  0.032  &  7, 16                   \\ 
 UGC 09614      &  7.23  & 8.18 $\pm$ 0.09 & -0.218 $\pm$ 0.128 (0.198) & 0.063 & 6.88 $\pm$ 0.14 & -0.389 $\pm$ 0.195 (0.262) &  0.096  &  7, 16                   \\ 
 UGC 12709      &  7.86  & 8.48 $\pm$ 0.08 & -0.358 $\pm$ 0.113 (0.136) & 0.084 & 7.31 $\pm$ 0.15 & -0.546 $\pm$ 0.210 (0.217) &  0.156  &  16                      \\ 
\hline
\end{tabular}\\
\end{center}
\begin{flushleft}
References: 
1 --  \citet{Berg2012}, 
2 -- \citet{Boker2001},
3 --  \citet{Croxall2009},
4 --  \citet{Egorov2013}, 
5 -- \citet{Esteban2009},  
6 --  \citet{GildePaz2000}, 
7 -- \citet{Haurberg2013},    
8 -- \citet{Ho1997}, 
9 --  \citet{Hunter1982}, 
10 -- \citet{Hunter1999}, 
11 --  \citet{Kobulnicky1996}, 
12 --  \citet{Lequeux1979},  
13 -- \citet{McCall1985}, 
14  -- \citet{Moustakas2006},  
15 --  \citet{Romanishin1983},   
16 --  SDSS (\citet{York2000},  
17 -- \citet{vanZee1998},  
18 -- \citet{vanZee2006}.  
\end{flushleft}
\end{table*}

The deprojected radii of the H\,{\sc ii} regions were computed using
their coordinates and geometrical parameters (position angle of the
major axis and galaxy inclination) listed in Table \ref{table:sample}.

The radial oxygen abundance distribution within the isophotal radius
in every galaxy was fitted by the following equation:
\begin{equation}
12+\log({\rm O/H})  = 12+\log({\rm O/H})_{R_{0}} + C_{O/H} \times (R/R_{25}) ,
\label{equation:grado}
\end{equation} 
where 12 + log(O/H)$_{R_{0}}$ is the oxygen abundance at $R_{0}$ = 0,
i.e., the extrapolated central oxygen abundance. C$_{O/H}$ is the
slope of the oxygen abundance gradient expressed in terms of
dex~$R_{\rm 25}^{-1}$, and $R$/$R_{\rm 25}$ is the fractional radius
(the galactocentric distance normalized to the disk's isophotal radius
$R_{25}$). The derived parameters of the oxygen abundance
distributions are presented in Table~\ref{table:grad}.  The name of
the galaxy is listed in column 1.  The optical isophotal radius
$R_{25}$ in kpc is reported in column 2.  The extrapolated central 12
+ log(O/H)$_{R_{0}}$ oxygen abundance and the gradient expressed in
terms of dex~$R_{\rm25}^{-1}$ are listed in columns 3 and 4 (the
bootstrapped error of the gradient is given in parenthesis). The
scatter of oxygen abundances around the general radial oxygen
abundance trend is reported in column 5.  The references to sources
for spectroscopic data are given in column 9 .  The radial
distributions of the oxygen abundances in irregular galaxies are shown
in Figs.~\ref{figure:ygrad} and \ref{figure:ngrad} together with the
surface brightness profiles. 

The statistical error of the gradient listed in column 4 comes
from the best fitting procedure. We also estimate the bootstrapped
error of the gradient in the following way.  The measured  H\,{\sc ii}
regions in a galaxy are numbered from 1 to $n$.  We then produce $n$
random integer numbers using a random number generator, and form a
bootstrapped subsample of H\,{\sc ii} regions choosing the
corresponding H\,{\sc ii} regions from the original sample of H\,{\sc
ii} regions.  The amount of H\,{\sc ii} regions in the bootstrapped
subsample is adopted to be equal to the amount of the H\,{\sc ii}
regions in the original sample.  Thus, some H\,{\sc ii} regions from
the original sample can be repeatedly included in the bootstrapped
subsample while other H\,{\sc ii} regions from the original sample
will not at all be included in the bootstrapped subsample. If a
bootstrapped subsample involves less than three different H\,{\sc ii}
regions then this subsample is rejected.  The abundance gradient for
the bootstrapped subsample is determined through the best fit, and the
error of the original gradient, i.e., the difference between the
values of the gradients for the bootstrapped subsample and for the
original sample of H\,{\sc ii} regions is obtained. We considered $k =
10^{5}$ bootstrapped subsamples and determined the bootstrapped error
of the gradient as  ($[(\sum difference^2_j)/k]^{1/2}$).  This
bootstrapped error of the oxygen abundance gradient is given in Table~\ref{table:grad},
 column 4 in parenthesis.  

The statistical and bootstrapped errors of the oxygen abundance
gradients are close to each other except in the case of the galaxy
UGC~2216 where the bootstrapped error exceeds dramatically the
statistical error.  This is caused by the following. The radial
abundance gradient in the UGC~2216 is strongly biased by an  H\,{\sc
ii} region at a galactocentric distance of 4.43 kpc.  When the
bootstrapped subsample does not contain this point then the value of
the radial abundance gradient is very uncertain since in this case the
gradient is determined from measurements at close galactocentric
distances. As a result, the bootstrapped error of the radial abundance
gradient for this galaxy is quite large.  

As in the case of the oxygen abundance, the radial nitrogen abundance 
distribution in every galaxy was fitted by the following equation:
\begin{equation}
12+\log({\rm N/H})  = 12+\log({\rm N/H})_{R_{0}} + C_{N/H} \times (R/R_{25}) . 
\label{equation:gradn}
\end{equation}
The derived parameters of the nitrogen abundance distributions are
presented in Table~\ref{table:grad}.  The extrapolated central 12 +
log(N/H)$_{R_{0}}$ nitrogen abundance and the gradient in terms of
dex~$R_{\rm 25}^{-1}$ are listed in columns 6 and 7.  The scatter of
oxygen abundances around the general radial oxygen abundance trend is
reported in column 8.  The value in the parenthesis in column 7
is the bootstrapped error of the nitrogen abundance gradient obtained
in the same way as for the oxygen abundance gradient.  

The radial oxygen abundance gradients in irregular galaxies obtained
here are based mainly (or only) on oxygen abundances (O/H)$_{C_{\rm
NS}}$ estimated through strong emission lines using the $C_{NS}$
method.  Figs.~\ref{figure:ygrad} and \ref{figure:ngrad} show that the
scatter in the (O/H)$_{C_{\rm NS}}$ abundances around the general
radial trend is often lower than the scatter in the (O/H)$_{T_{e}}$
abundances.  Five galaxies from our present sample are in the
list of galaxies considered in our previous study
\citep{Pilyugin2014a}. The values of gradients obtained here are
slightly different from those reported in our previous study for the
following reasons. First, in our current work we obtain and use new
parameters for our target galaxies such as inclination, position angle
of the major axis, and optical isophotal radius. Furthermore, in
\citet{Pilyugin2014a} the oxygen and nitrogen abundances were
estimated via the  $C_{ON}$ method for H\,{\sc ii} regions with
available measurements of the [O\,{\sc ii}]$\lambda$$\lambda$3727,3729
emission line, and with the  $C_{NS}$ method for the other H\,{\sc ii}
regions.   In our current study, the oxygen and nitrogen abundances
were estimated through the  $C_{NS}$ method for all H\,{\sc ii}
regions, and $T_{e}$-based abundances are added.

\section{Discussion and conclusions}

\begin{figure*}
\resizebox{1.00\hsize}{!}{\includegraphics[angle=000]{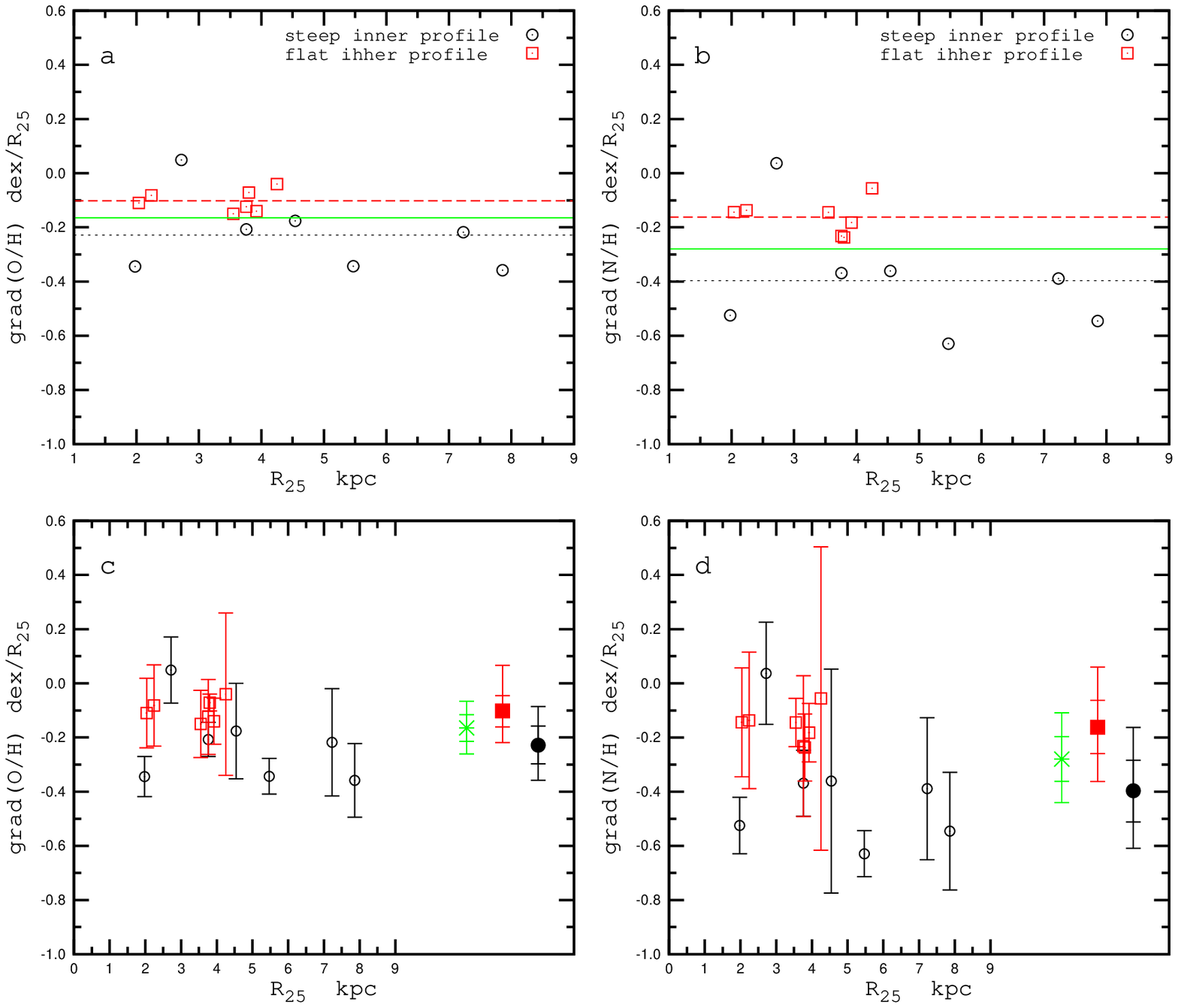}}
\caption{  
The panel $a$ shows the oxygen abundance gradient as a function of optical radius $R_{25}$.  
The dark (black) open circles mark irregular galaxies with steep inner photometric profiles.  
The dark-grey (red) open squares denote galaxies with flat inner photometric profiles.  
The dark (black) dotted line is the arithmetic mean of the gradients for galaxies
with steep inner profiles, the dark-grey (red) dashed line for galaxies with flat inner profiles, 
and the light-grey (green) solid line  the total sample.
Panel $b$ shows the same as panel $a$ but for the nitrogen abundance
gradients.  Panel $c$ shows the oxygen abundance gradients with
bootstrapped errors.  On the right side of the panel, the mean values
of the gradients for the sample of galaxies with steep inner
photometric profiles (the filled dark (black) circle), for the sample
of galaxies with flat inner photometric profiles (the filled dark-grey
(red) square), and for total sample (the ligh-grey (green) asterisk)
and their 95\% and 68\% confidence intervals are shown.  Panel $d$ shows the
same as panel $c$ but for the nitrogen abundance gradients.  (A color
version of this figure is available in the on-line edition.)
}
\label{figure:r25grad}
\end{figure*}

Fig.~\ref{figure:ygrad} shows the surface brightness profiles and
radial distributions of the oxygen abundances for the irregular
galaxies with steep inner profiles. Each galaxy is presented in two
panels. Each upper panel $N$a shows the surface brightness profiles in
the SDSS $g$ band as a light-grey (blue) solid line, in the SDSS $r$
band as a dark (black) long-dashed line, and in the WISE $W1$ band as
a dark-grey (red) short-dashed line.  Each lower panel $N$b shows the
oxygen abundance in individual H\,{\sc ii} regions (open circles) as a
function of radius.  The linear best fit to those data is indicated by
a solid line.  Fig.~\ref{figure:ygrad} shows that irregular galaxies
with steep inner profiles have appreciable radial abundance
gradients.  

Fig.~\ref{figure:ngrad} shows the surface brightness profiles and
radial distributions of the oxygen abundances for irregular galaxies
with flat inner profiles.  Inspection of Fig.~\ref{figure:ngrad} shows
that the radial abundance gradients in the irregular galaxies with
flat inner profiles are shallower than the gradients in irregular
galaxies with steep inner profiles. 

Thus, our data suggest that there is a relation between the radial
abundance gradient in an irregular galaxy and its surface brightness
profile.  Panel $a$ of Fig.~\ref{figure:r25grad} shows the radial
oxygen abundance gradient as a function of optical radius $R_{25}$ for
our sample of irregular galaxies.  The dark (black) open circles mark
irregular galaxies with steep inner photometric profiles.  The
dark-grey (red) open squares denote galaxies with flat inner profiles.
The dark (black) dotted line is the arithmetic mean of the gradients
for galaxies with steep inner photometric profiles, whereas the
dark-grey (red) dashed line is the mean for galaxies with flat inner
photometric profiles. The light-grey (green) solid line is the
arithmetic mean of the gradients for all our galaxies (both those with
steep and those with flat inner profiles).  Since the numbers of
galaxies in our samples are small even one deviant galaxy may
appreciably change the arithmetic mean for the sample.  Indeed the
aritmetic mean of the gradients for galaxies with steep inner
photometric profiles is changed by $\sim$0.05 dex $R_{25}^{-1}$ when
the deviating galaxy NGC~4214 (with a positive gradient 0.049 dex
$R_{25}^{-1}$) is excluded from consideration.  

Panel $b$ of  Fig.~\ref{figure:r25grad} shows the same as
panel $a$ but for the nitrogen abundance gradients. Comparison between
panels $a$ and $b$ shows that the general picture is similar for
oxygen and nitrogen abundance gradients, i.e., the irregular galaxies
with flat inner photometric profiles have shallower nitrogen abundance
gradients as compared to galaxies with steep inner photometric
profiles.  

Panel $c$ of Fig.~\ref{figure:r25grad} shows the radial oxygen
abundance gradients with bootstrapped errors (reported in
Table~\ref{table:grad} in parenthesis).  The filled dark (black)
circle on the right side of the panel shows the mean value of the
gradients within its 95\% and 68\% confidence intervals for the sample of
galaxies with steep inner photometric profiles.  To estimate the
confidence interval of the mean value of the gradients of the sample
of galaxies the bootstrap method is used.  We create $10^{5}$
bootstrapped subsamples from the original sample of gradients keeping
the size of each bootstrapped subsample equal to the size of the
original sample, and modifying the value of the original gradient of
each galaxy by introducing a random error. This error is randomly
chosen from a set of errors that follow a Gaussian distribution scaled
to the standard deviation corresponding to the bootstrapped error of
abundance gradient (reported in  Table~\ref{table:grad} in
parenthesis).  We consider the distribution of the mean values of the
abundance gradients for those 10$^{5}$ bootstrapped subsamples and
determine the 95\% and 68\% confidence intervals of the mean abundance gradient
for the sample of galaxies.  The filled dark-grey (red) square shows
such a mean value of the abundance gradients for the sample of
galaxies with flat inner photometric profiles, and the light-grey
(green) asterisk shows the one for the total sample of galaxies.
Panel $d$ of Fig.~\ref{figure:r25grad} shows the same as panel $c$ but
for the radial nitrogen abundance gradients.   

The difference between the mean values of the oxygen abundance
gradients for galaxies with steep and flat inner photometric profiles
is estimated in a similar way and amounts to $-0.126$ dex
$R_{25}^{-1}$ within the 95\% confidence interval ($-0.306$, 0.059).
The difference between the mean values of the nitrogen abundance
gradients is $-0.236$ dex $R_{25}^{-1}$ within the 95\% confidence
interval ($-0.540$, 0.073).  The difference between the mean values of
the abundance gradients in irregular galaxies with steep and flat
inner photometric profiles exists (is less than 0) at 91\% confidence
level for oxygen abundance gradients and at 94\% confidence level for
nitrogen abundance gradients.  

Thus, our data suggest that {\it i)} there are radial abundance
gradients in irregular galaxies, and {\it ii)} there is a difference
between radial abundance gradients in irregular galaxies with steep
and flat inner photometric profiles with a probability higher than
90\%.  

It should be noted that here the abundances are determined
through the $C$ and $T_{e}$ methods.  The $C$ method is based on the
abundances derived via the $T_e$ method and, consequently, produces
the abundances on the same metallicity scale as the $T_e$ method.  If
the abundances derived using the $T_e$ method are not correct for some
reason (e.g., because of small-scale temperature fluctuations within
an H\,{\sc ii} region \citep{Peimbert1967}, or if the energies of the
electrons in an H\,{\sc ii} region do not follow a Maxwell
distribution \citep{Dopita2013}) then our abundances should be
revised.  Furthermore, the absolute metallicity scale of  H\,{\sc ii}
regions varies up to $\sim$0.7 dex depending on the calibration used
\citep{Kewley2008}.  As was noted above, a prominent characteristic
of the previous calibrations is that they are not applicable across
the whole range of metallicities of H\,{\sc ii} regions but only
within a limited interval.  The oxygen abundances of irregular
galaxies typically are within or near the transition zone in the
$R_{23}$ -- O/H diagram where previous calibrations cannot be used or
where they provide abundances with large uncertainties.  Therefore the
$T_e$- and $C$-based abundances are preferable for irregular galaxies.

It is known \citep[e.g.,][]{Searle1972,Pagel1997} that the radial
distribution of oxygen abundances in the disk of a galaxy is
controlled by the variation of the astration level (or gas mass
fraction $\mu$) with radius and by the mass exchange between a galaxy
and the surrounding medium (via galactic winds and/or gas
infall/merging) and between different parts of a galaxy.  Taking into
consideration the radial variation of the astration level, one may
expect that physical gradients (expressed in dex kpc$^{-1}$) in
irregular galaxies can be even steeper than those in spiral galaxies.
The metallicities in irregular galaxies are typically lower than the
ones in spiral galaxies since irregular galaxies are less massive and
less evolved.  The simple model for the chemical evolution of galaxies
predicts that the oxygen abundance O/H varies with gas mass fraction
$\mu$ more strongly at low metallicity. Thus a similar change of $\mu$
along the radial direction would result in a larger change of O/H in
irregular galaxies than in spiral galaxies. 

Radial mixing of gas flattens the abundance gradient in the disk
of a galaxy. Radial mixing of gas can be caused by interacting or
merging galaxies \citep[e.g.,][]{Rupke2010a,Rupke2010b} and by
galactic fountains (galactic winds and subsequent gas infall).  The
arguments pro and contra galactic wind-dominated evolution of
irregular galaxies are discussed in many studies devoted to the
chemical evolution of galaxies \citep[][among many
others]{Skillman1997,Cavilan2013}.  A galactic wind can be caused by
the injection of energy by multiple, spatially and temporally
clustered supernovae in a galaxy undergoing a starburst
\citep{DeYoung1990,MacLow1999}.  The efficiency of the galactic winds
depends on the number of massive stars that are progenitors of
supernovae in a star formation event.  \citet{Lee2009} found that
continuous, steady star formation dominates in the present epoch in
dwarf galaxies. Only $\sim6$\% of low-mass galaxies experience strong
star formation bursts. The fraction of stars formed in starbursts is
$\sim23$\%. However, it is not clear whether a strong star formation
burst can occur with equal probability in every galaxy or whether a
starburst happens only in a particular subset of galaxies.  

Thus, we can interpret our results in the following manner.  Irregular
galaxies with steep inner profiles do not seem to undergo strong
radial mixing of gas at the present epoch and show considerable radial
abundance gradients.  The radial mixing of gas (through radial flows
or galactic fountains) took place in irregular
galaxies with flat inner profiles, resulting in shallower (if any)
gradients as compared to the galaxies with steep inner profiles.  It
should be noted that the physical reason for different radial profile
types is still a mystery.  It is not even clear why there is an
exponential drop-off of the brightness profile \citep{Herrmann2013}.

\section {Summary}

We determined the abundance distributions traced by H\,{\sc ii} regions
and compare their shape with the surface brightness profiles of the
disks of fourteen irregular {\em Sm} and {\em Im} galaxies
(morphological $T$ types of $T$ = 9 and $T$ =10).  We used the emission
line intensities in published spectra of  H\,{\sc ii} regions from
different studies to infer the abundances. The oxygen (O/H)$_{T_{e}}$
and nitrogen (N/H)$_{T_{e}}$ abundances in the H\,{\sc ii} regions
with the detected auroral line [O\,{\sc iii}]$\lambda$4363 were
determined using the equations of the classic $T_{e}$-method.  In the
other H\,{\sc ii} regions, oxygen (O/H)$_{C_{\rm NS}}$ and nitrogen
(N/H)$_{C_{\rm NS}}$ abundances were obtained through the $C$ method.
We then quantified the values of the gradients of the radial abundance
profiles.  

Moreover, we constructed radial surface brightness profiles in the
infrared $W1$ WISE band and in the SDSS $g$ and $r$ bands using the
publicly available photometric maps.  The irregular galaxies of our
sample can be divided into two types according to the shapes of their
surface brightness profiles: those with steep inner profiles, and
those with flat inner profiles. 

We find that there is a correspondence between the radial abundance
gradient in an irregular galaxy and its surface brightness profile
with a probability higher than 90\%.  Irregular galaxies with
steep inner profiles usually show a considerable radial abundance
gradient.  Irregular galaxies with flat inner surface brightness
profiles have shallower gradients (if any) as compared to galaxies
with steep inner profiles. 

Thus, irregular galaxies with steep inner profiles show usually a
pronounced radial abundance gradient that resembles that of spiral
galaxies.  In that sense, those irregular galaxies seem to extend the
Hubble sequence of spiral galaxies.  In other words, our data suggest
that there is no ``spiral versus irregular dichotomy'' in terms of
radial abundance gradients existing only in spiral galaxies, but not
in irregulars.  While irregulars have long been believed to be
chemically homogeneous, our study shows that given enough measurements
of nebular abundances of H\,{\sc ii} regions across a wide range of
galactocentric radii, irregulars may well exhibit radial abundance
gradients.  This tendency is particularly conspicuous in irregulars
with steep surface brightness profiles in their inner regions.

\section*{Acknowledgements}

We are grateful to the referee for his/her constructive comments. \\
L.S.P., E.K.G., and I.A.Z.\  acknowledge support within the framework
of Sonderforschungsbereich (SFB 881) on "The Milky Way System"
(especially subproject A5), which is funded by the German Research
Foundation (DFG). \\ L.S.P. and I.A.Z thank the hospitality of the
Astronomisches Rechen-Institut at Heidelberg University  where part of
this investigation was carried out. \\
This work was partly funded by the subsidy allocated to Kazan Federal 
University for the state assignment in the sphere of scientific 
activities (L.S.P.).  \\ 
We thank  R.A.~Swaters and
M.~Balcells for supporting us with the surface brightnes profiles of
galaxies from their sample in numerical form. \\ This research made
use of Montage, funded by the National Aeronautics and Space
Administration's Earth Science Technology Office, Computational
Technnologies Project, under Cooperative Agreement Number NCC5-626
between NASA and the California Institute of Technology. The code is
maintained by the NASA/IPAC Infrared Science Archive. \\ Funding for
the SDSS and SDSS-II has been provided by the Alfred P. Sloan
Foundation, the Participating Institutions, the National Science
Foundation, the U.S. Department of Energy, the National Aeronautics
and Space Administration, the Japanese Monbukagakusho, the Max Planck
Society, and the Higher Education Funding Council for England. The
SDSS Web Site is http://www.sdss.org/.

\end{document}